\documentstyle[aps,preprint,prc]{revtex}
\begin{document}
\draft
\title{\bf Baryon Mapping of Quark Systems}
\author{M. Sambataro}
\address{Istituto Nazionale di Fisica Nucleare - Sezione di Catania;\\
Corso Italia, 57 - I 95129 Catania - Italy}

\maketitle
\begin{abstract}
We discuss a mapping procedure from a space of colorless
three-quark clusters into a space of elementary baryons
and illustrate it in the context of
a three-color extension of the Lipkin model recently developed.
Special attention is addressed to the
problem of the formation of unphysical states in the mapped space.
A correspondence  is established between quark and baryon spaces
and the baryon image of a generic quark operator is defined both in
its Hermitian and non-Hermitian forms. Its spectrum (identical in the
two cases) is found to consist of a physical part containing the
same eigenvalues of the quark operator in the cluster space and an
unphysical part consisting only of zero eigenvalues. A physical
subspace of the baryon space is also defined where the latter eigenvalues
are suppressed. The procedure discussed is quite general and applications
of it can be thought also in the correspondence between systems of 2n
fermions and n bosons.
\newline
\  \\
\  \\
PACS number(s): 24.85.+p, 12.39.-x, 13.75.cs, 21.45.+v
\end{abstract}

\newpage
\section*{1. Introduction}

Among the QCD-inspired quark models of baryons, non-relativistic
constituent models
have attracted considerable attention in recent years \cite{isgur}.
Here, baryons are
assumed to be clusters of three quarks, each of them carrying
color, spin and isospin and interacting via a potential whose main terms
are a confining and a hyperfine term. These models have
provided interesting results in the description of single baryon properties
and, based on that, attempts have been made to extend their application to
the study of the baryon-baryon interaction \cite{shimi} as well as of
few-baryon bound systems \cite{yama}.

It is in this framework that mapping techniques traditionally
developed in nuclear physics for the description of collective excitations
and establishing a correspondence between systems of 2n fermions and n
bosons \cite{klein} have been recently extended to the correspondence
between systems of 3n fermions and n fermions.
More precisely, it is the mapping of three-quark clusters
onto ``elementary baryons'', namely fermions carrying the same
quantum numbers as the clusters, which has become the object of investigation.

Recent works on this subject have been those of Nadjakov in 1990
\cite{nadja}, Pittel, Engel, Dukelsky and Ring in 1990 \cite{pedr}
and Meyer in 1991 \cite{meyer}. Although different among
themselves, these procedures all have a common point: they follow the
Belyaev-Zelevinsky method which is that based on the mapping of the
operators in such a way that their
commutation relations are preserved \cite{bely}.
More particularly, it is the Dyson mapping \cite{dyson} or generalizations
of it which they employ. Applications of these procedures
can be found within the
so-called Quark Nuclear-Plasma Model of Nadjakov \cite{nadja} as well as
within the so-called Bonn Quark Shell Model of Petry et al. \cite{petry},
in Ref.(7).

A quite interesting scenario appeared in the more recent
paper (1994) of Pittel, Arias, Dukelsky and Frank \cite{padf} (hereafter
referred to as PADF). Here, the authors have developed
 a three-color extension of the
so-called Lipkin-Meshkov-Glick model \cite{lipkin} which has been widely used
in the past as a testing bench for nuclear many-body approximations. The
quark Hamiltonian of the model
includes one-body, two-body and three-body interactions
and, as for the model in its original form,
group theoretical techniques
have been developed for an exact solution of its eigenvalue problem.

By following also in this case the Belyaev-Zelevinsky method,
PADF have developed a new mapping procedure which
has been found able to
overcome some limitations evidenced in the previous approaches
\cite{padf}. Among these limitations, for instance, the ``preference''
of these approaches toward special forms of Hamiltonians.
The image of the Lipkin Hamiltonian has been constructed by PADF in both
a Hermitian and a non-Hermitian form and, in both cases, all the
original quark eigenvalues have been exactly reproduced in the baryon space.
However, besides these eigenstates,
all with a corresponding one in the quark cluster space, several
other states have appeared which are a pure artifact of the mapping procedure.
It is the mixing in the spectrum of these ``physical''
and ``unphysical'' states which has been analyzed by PADF.

In 1991, Catara and Sambataro \cite{cs} (hereafter referred to as CS) have
proposed a mapping procedure which is different from those discussed so far
in that it does not follow the Belyaev-Zelevinsky method. The starting point
of the procedure has been a ``simple'' (as  will
also be discussed later) correspondence between a space of quark clusters
and a space of elementary baryons. Therefore the baryon image of a
generic quark operator has been constructed
such that all the eigenvalues of the quark operator in the
cluster space were also eigenvalues of its image.
This does not imply that corresponding matrix elements of the
quark operator in the cluster space and of its image  must be equal.
However, a further correspondence has also been established between
quark and baryon spaces
such that matrix elements were indeed preserved as within the
so-called Marumori approach \cite{maru}.

This procedure has been first applied to a realistic Hamiltonian of Oka
and Yazaki \cite{oka} and the derived nucleon-nucleon Hamiltonian analyzed
\cite{cs}. As a second application \cite{cs2}, the authors have derived
the nucleon image of the one-body quark density operator and expectation
values of this operator have been calculated in the ground state of doubly
magic nuclei like $^4$He, $^{16}$O and $^{40}$Ca described within the nuclear
shell model. This has allowed an analysis of quark exchange effects on the
quark densities of these nuclei.

As also evidenced by CS, the realizations just discussed have referred to cases
in which corresponding states were forming a set of linearly
independent states on one side, the {\it composite} space, and a set of
orthonormal states on the other side, the {\it elementary} space.
This has created the conditions for
the non-appearance of unphysical states in the
mapped space. In circumstances different from these, the appearance of
these states would have made the mapping considerably more complicated
and a study of this problem
was left to future developments of the theory.

The three-color extension of the Lipkin model proposed by PADF has offered the
opportunity of investigating this problem thoroughly. The mapping procedure
of CS has now been reviewed with reference to the new model.
After establishing a ``simple'' correspondence (as in CS) between the spaces of
three-quark clusters and of elementary baryons,
the baryon image of a quark operator has been defined, in both its Hermitian
and non-Hermitian forms, and its spectrum analyzed.
As a general result, this spectrum (identical in the two cases)
has been found to consist of (a) a physical part whose eigenvalues
are identical to those of the quark operator in the cluster space
and (b) an unphysical part
whose eigenvalues are all zero. Moreover,
a further correspondence between quark and baryon spaces
has been established
such as to guarantee the equality of corresponding matrix elements
and so a physical baryon subspace has been defined.

As an important point, we remark that
the procedure which is discussed in this paper has been
developed in a quite general form so
that applications of it can be considered for very different cases like, for
instance, the correspondence between systems of 2n fermions and n bosons.

The paper is organized as follows. In sect.2, we briefly review the three-color
extension of the Lipkin model developed by PADF. In sect.3, we discuss the
mapping procedure, and precisely: in subsect. 3.1, we establish the
correspondence between the quark and baryon spaces; in subsect. 3.2, we
derive the baryon image of a quark operator in its non-Hermitian form;
in subsect. 3.3, we derive the image in its Hermitian form.
In sect.4, we discuss the n-body structure of the image operator and
consider, as an example, the Hamiltonian of the Lipkin model.
Finally, in sect.5, we summarize the results and give some closing remarks.

\section*{2. The three-color Lipkin model.}

As anticipated in the introduction, the three-color Lipkin model has
been presented and discussed thoroughly by PADF. Here, we will briefly review
its main points.

The model is a natural extension of the standard Lipkin model \cite{lipkin} to
fermions characterized by three colors. Therefore, there are two levels,
each one $3\Omega$-fold degenerate, separated by an energy $\Delta$. Each
single-particle state in these levels is characterized by three quantum
numbers: $c$, the color, $\sigma$, which individuates whether
the state belongs to the level ``up'' ($\sigma =+$) or ``down'' ($\sigma =-$)
and, finally, $p$, which runs from 1 up to $\Omega$. In the
unperturbed ground state, it is assumed that $N=3\Omega$ particles
occupy all the single-particle states in the lower level.

The model is discussed in second quantized form and so creation
and annihilation operators $q^{\dag}_{c\sigma p}$ and $q_{c\sigma p}$
are introduced. These satisfy the fermion commutation relations
\begin{equation}
\{q^{\dag}_{c\sigma p},q^{\dag}_{c'\sigma 'p'}\}=
\{q_{c\sigma p},q_{c'\sigma 'p'}\}=0\label{2.1}~,
\end{equation}
\begin{equation}
\{q_{c\sigma p},q^{\dag}_{c'\sigma 'p'}\}=\delta_{c,c'}\delta_{\sigma,\sigma '}
\delta_{p,p'}\label{2.2}
\end{equation}
where $\{A,B\}=AB+BA$.
The model Hamiltonian includes one-body, two-body and three-body interactions
and scatters particles among the levels without changing the $p$ values
and maintaining all states ``colorless'' (as will be pointed
out in the next section). Its form is
\begin{equation}
{\hat H}_{C}={\hat H}^{(1)}_{C}+{\hat H}^{(2)}_{C}+{\hat H}^{(3)}_{C}
\label{2.3}~,
\end{equation}
with
\begin{equation}
{\hat H}^{(1)}_{C}=\slantfrac{\Delta}{2}\sum_{cp}(q^{\dag}_{c+p}q_{c+p}-
q^{\dag}_{c-p}q_{c-p})\label{2.4}~,
\end{equation}
\begin{equation}
{\hat H}^{(2)}_{C}=-\slantfrac{\chi_2}{\Omega}\sum_{c_1c_2c_3c_4c_5p_1p_2}
\epsilon_{c_1c_2c_3}\epsilon_{c_1c_4c_5}
(q^{\dag}_{c_2+p_1}q^{\dag}_{c_3+p_2}q_{c_5-p_2}q_{c_4-p_1}+
q^{\dag}_{c_4-p_1}q^{\dag}_{c_5-p_2}q_{c_3+p_2}q_{c_2+p_1})\label{2.5}~
\end{equation}
and
\begin{eqnarray}
{\hat H}^{(3)}_{C}=-\slantfrac{\chi_3}{{\Omega}^2}
\sum_{c_1c_2c_3c_4c_5c_6p_1p_2p_3}
\epsilon_{c_1c_2c_3}\epsilon_{c_4c_5c_6}
&&(q^{\dag}_{c_1+p_1}q^{\dag}_{c_2+p_2}q^{\dag}_{c_3+p_3}
q_{c_6-p_3}q_{c_5-p_2}q_{c_4-p_1}\nonumber\\
&&+q^{\dag}_{c_4-p_1}q^{\dag}_{c_5-p_2}q^{\dag}_{c_6-p_3}
q_{c_3+p_3}q_{c_2+p_2}q_{c_1+p_1})
\label{2.6}~,
\end{eqnarray}
where $\epsilon_{c_1c_2c_3}$ is the totally antisymmetric tensor of rank 3.

\section*{3. Baryon mapping of quark operators.}
\subsection*{3.1 Quark and baryon spaces: the correspondence}
Let us define
\begin{equation}
F^{\dag}_{\mu_1\mu_2\mu_3}=\slantfrac{1}{6}\sum_{c_1c_2c_3}\epsilon_{c_1c_2c_3}
q^{\dag}_{c_1\mu_1}q^{\dag}_{c_2\mu_2}q^{\dag}_{c_3\mu_3}\label{3.11}
\end{equation}
(where $q^{\dag}_{c\mu}\equiv q^{\dag}_{c\sigma p}$) the operator which
creates a colorless cluster of three particles (the ``quarks'')
characterized by the
quantum numbers $\sigma_1p_1, \sigma_2p_2, \sigma_3p_3$.
This operator is symmetric with respect to the indices $1, 2, 3$.
We define $C^{(\Omega )}$ the vector space
spanned by the states which are obtained by acting with
$\Omega$ cluster creation operators on a vacuum state $|0\rangle$,
\begin{equation}
F^{\dag}_{\mu^{(1)}_1\mu^{(1)}_2\mu^{(1)}_3}
F^{\dag}_{\mu^{(2)}_1\mu^{(2)}_2\mu^{(2)}_3}...
F^{\dag}_{\mu^{(\Omega)}_1\mu^{(\Omega)}_2\mu^{(\Omega)}_3}|0\rangle
\label{3.12}~,
\end{equation}
the vacuum being defined by the condition
\begin{equation}
q_{c\mu}|0\rangle=0\label{3.13}~.
\end{equation}
In the following, we will discuss in some detail the cases $\Omega=1$ and
$2$. The second one is particularly interesting since the associated mapping
is representative of all cases with larger $\Omega$.

States of $C^{(1)}$ are
\begin{equation}
|\mu_1\mu_2\mu_3\rangle\equiv F^{\dag}_{\mu_1\mu_2\mu_3}|0\rangle
\label{3.14}
\end{equation}
and their overlap is
\begin{equation}
\langle \mu_1\mu_2\mu_3|\mu'_1\mu'_2\mu'_3\rangle=
\slantfrac{1}{6}S(\mu_1\mu_2\mu_3, \mu'_1\mu'_2\mu'_3)
\label{3.15}
\end{equation}
where
\begin{equation}
S(\mu_1\mu_2\mu_3, \mu'_1\mu'_2\mu'_3)=\sum^{3}_{i,j,k=1}|\epsilon_{ijk}|
\delta_{\mu_1,\mu'_i}\delta_{\mu_2,\mu'_j}\delta_{\mu_3,\mu'_k}
\label{3.16}~.
\end{equation}
States of $C^{(2)}$ are
\begin{equation}
|i\rangle=F^{\dag}_{\mu^{(i)}_1\mu^{(i)}_2\mu^{(i)}_3}
F^{\dag}_{\mu^{(i)}_{1'}\mu^{(i)}_{2'}\mu^{(i)}_{3'}}|0\rangle
\label{3.17}
\end{equation}
and, differently from the previous case, they are neither orthogonal
nor linearly independent. By constructing
the overlap matrix and diagonalizing
it, one finds indeed ${\overline N}=20$ orthonormal states to be compared
with the total number of $N=52$ states.

In correspondence with the quark cluster operator
$F^{\dag}_{\mu_1\mu_2\mu_3}$ (\ref{3.11}), let us define the ``baryon''
operator
$f^{\dag}_{\mu_1\mu_2\mu_3}$, symmetric with respect to the three indices
and satisfying the commutation relations
\begin{equation}
\{f^{\dag}_{\mu_1\mu_2\mu_3}, f^{\dag}_{\mu'_1\mu'_2\mu'_3}\}=
\{f_{\mu_1\mu_2\mu_3}, f_{\mu'_1\mu'_2\mu'_3}\}=0
\label{3.18}
\end{equation}
\begin{equation}
\{f_{\mu_1\mu_2\mu_3}, f^{\dag}_{\mu'_1\mu'_2\mu'_3}\}=
\slantfrac{1}{6}S(\mu_1\mu_2\mu_3, \mu'_1\mu'_2\mu'_3)
\label{3.19}~,
\end{equation}
where the function $S(\mu_1\mu_2\mu_3, \mu'_1\mu'_2\mu'_3)$ is
defined in Eq.(\ref{3.16}). We call $E^{(\Omega)}$ the vector space spanned by
the states which are generated by the action of $\Omega$ of these
operators on a vacuum $|0)$
\begin{equation}
f^{\dag}_{\mu^{(1)}_1\mu^{(1)}_2\mu^{(1)}_3}
f^{\dag}_{\mu^{(2)}_1\mu^{(2)}_2\mu^{(2)}_3}...
f^{\dag}_{\mu^{(\Omega)}_1\mu^{(\Omega)}_2\mu^{(\Omega)}_3}|0)
\label{3.20}~.
\end{equation}
A ``simple'' (see also \cite{cs}) correspondence can be established between the
states of $C^{(\Omega )}$ and $E^{(\Omega )}$. States (\ref{3.20}) can be,
in fact,
formally obtained from states (\ref{3.12}) by simply replacing cluster creation
operators $F^{\dag}_{\mu^{(i)}_1\mu^{(i)}_2\mu^{(i)}_3}$ with baryon
creation operators $f^{\dag}_{\mu^{(i)}_1\mu^{(i)}_2\mu^{(i)}_3}$ and the
quark vacuum $|0\rangle$ with the baryon vacuum $|0)$. In correspondence
with the state (\ref{3.14}) of $C^{(1)}$, for instance, we have for $E^{(1)}$
\begin{equation}
|\mu_1\mu_2\mu_3)\equiv f^{\dag}_{\mu_1\mu_2\mu_3}|0)
\label{3.21}
\end{equation}
and, similarly, in correspondence with the state (\ref{3.17}) of $C^{(2)}$,
we have for $E^{(2)}$
\begin{equation}
|i)=f^{\dag}_{\mu^{(i)}_1\mu^{(i)}_2\mu^{(i)}_3}
f^{\dag}_{\mu^{(i)}_{1'}\mu^{(i)}_{2'}\mu^{(i)}_{3'}}|0)
\label{3.22}~.
\end{equation}
An important feature of these states is their orthogonality. This
orthogonality,
on one side, and the linear dependence of the corresponding states of
$C^{(\Omega )}$ (at least for $\Omega >1$), on the other side, clearly
reflect the ``elementary'' or ``composite'' nature of the baryons entering
in the definitions of the spaces $E^{(\Omega )}$ and $C^{(\Omega )}$,
respectively, and are associated with the different commutator algebras
of the operators $f^{\dag}_{\mu_1\mu_2\mu_3}$ and $F^{\dag}_{\mu_1\mu_2\mu_3}$.

Results similar to those just discussed in the correspondence between spaces
$C^{(\Omega )}$ and $E^{(\Omega )}$ for $\Omega =2$ also hold for larger spaces
$(\Omega =3,4,...)$. Also in these cases, in fact, correspondent states while
forming an orthogonal set in $E^{(\Omega )}$ are a set of linearly dependent
states in $C^{(\Omega )}$.

In the next section, we will describe a procedure
aiming to derive the image in $E^{(\Omega )}$ of a generic operator acting
within $C^{(\Omega )}$. For simplicity, we will refer to the case $\Omega =2$
and, therefore, to the correspondence between
the states (\ref{3.17}) and (\ref{3.22}). However, the value of $\Omega$
having no relevance  (only $\Omega >1$), if not
for the difficulty in actual calculations, in the following
we will suppress the indication of $\Omega$.

\subsection*{3.2 The image operator and its physical and unphysical
eigenstates: its non-Hermitian form}

On the basis of what has been said at the end of the last subsection, let
$C$ be the vector space spanned by the $N$ states
$\{|1\rangle, |2\rangle,... , |N\rangle\}$ defined by Eq.(\ref{3.17}).
Similarly, let $E$ be the vector space spanned by the $N$ states
$\{|1), |2),... , |N)\}$ defined by Eq.(\ref{3.22}). With respect to this
last equation, we are only supposing
that these states have been normalized so that they now satisfy the
condition
\begin{equation}
(i|j)=\delta_{i,j}~~~~~~~~~\forall i,j=1,2,...,N
\label{3}~.
\end{equation}
In this section, in correspondence with a Hermitian operator ${\hat O}_C$
acting
within $C$, we will search for an operator ${\hat O}_E$ acting
within $E$ such that all the eigenvalues of ${\hat O}_C$ in $C$ be also
eigenvalues of ${\hat O}_E$ in $E$. We will refer to ${\hat O}_E$
as the image operator of ${\hat O}_C$ in $E$. This operator will be first
derived in its non-Hermitian form ${\hat O}^{(nH)}_E$.
The study of the Hermitian image
${\hat O}^{(H)}_E$ will then be reserved to the next subsection.

Let ${\overline N}$ be the number of orthonormal states which can be
constructed
in terms of the $N$ states $|i\rangle$: therefore, the space $C$ is
${\overline N}$-dimensional. In subsect.(3.1) we have already
seen that it is $N=52$
and ${\overline N}=20$ in the case $\Omega =2$. In all coming equations
of this subsection as well as of the next one, indices written in terms of
the latin letters $k,i,j,...$ will be
meant to vary in the interval $(1,N)$ while
those written in terms of the greek letters $\alpha,\beta,\gamma,...$ will be
meant to vary in the interval $(1,{\overline N})$.

Each orthonormal state within $C$, $\overline{|\alpha\rangle}$, is a linear
combination of states $|i\rangle$ which we write as
\begin{equation}
\overline{|\alpha\rangle}=\slantfrac{1}{\sqrt{{\cal N}_\alpha}}\sum_{i}
f_{i\alpha}|i\rangle\equiv \sum_{i}{\overline f}_{i\alpha}|i\rangle ~,
\label{4}
\end{equation}
where $f_{i\alpha}$ and ${\cal N}_\alpha$ are real quantities
satisfying the equations
\begin{equation}
\sum_{l}\langle i|l\rangle f_{lj}={\cal N}_{j}f_{ij}
\label{5}~,
\end{equation}
\begin{equation}
\sum_{i}f_{ij}f_{ij'}=\delta_{jj'}
\label{6}
\end{equation}
and
\begin{equation}
\sum_{j}f_{ij}f_{i'j}=\delta_{ii'}
\label{7}~.
\end{equation}
The identity operator within $C$ is
\begin{equation}
{\hat I}_C=\sum_{\alpha}\overline{|\alpha}\rangle\langle\overline{\alpha|}
\label{8}
\end{equation}
and, by making use of Eq.(\ref{4}), it can be written as
\begin{equation}
{\hat I}_C=\sum_{ij}|i\rangle B(i,j)\langle j|
\label{9}
\end{equation}
with
\begin{equation}
B(i,j)=\sum_{\alpha}{\overline f}_{i\alpha}{\overline f}_{j\alpha}
\label{10}~.
\end{equation}
Only in the special case of a set of linearly independent
states $(|1\rangle , |2\rangle ,... , |N\rangle )$
(i.e., ${\overline N}=N$) would this matrix coincide with the
inverse of the overlap matrix $\langle i|j\rangle$.

We observe that, in general,
\begin{equation}
{\hat O}_C|i\rangle\not\in C
\label{100}~.
\end{equation}
By defining the operator ${\hat{\overline O}}_C\equiv{\hat I}_C{\hat O}_C$,
we notice that
\begin{equation}
(i)~~~{\hat{\overline O}}_C|l\rangle=\sum_i|i\rangle\{\sum_jB(i,j)
\langle j|{\hat O}_C|l\rangle\}
\label{101}~,
\end{equation}
\begin{equation}
(ii)~~~\langle i|{\hat{\overline O}}_C|l\rangle=\langle i|{\hat O}_C|l\rangle
\label{102}~.
\end{equation}
Namely, $(i)$ ${\hat{\overline O}}_C$ still gives rise to a state of $C$
when acting on a state of this space and $(ii)$ ${\hat{\overline O}}_C$
is equivalent to ${\hat O}_C$ within $C$.

To understand the role played by the operator ${\hat{\overline O}}_C$, let us
notice that the
eigenvalues $\lambda_\gamma$ of an operator ${\hat O}_C$, within the space
$C$, can be found by solving the system of ${\overline N}$ equations
\begin{equation}
\sum_{\beta}\overline{\langle\alpha|}{\hat
O}_C\overline{|\beta\rangle}c_{\beta\gamma}=
\lambda_{\gamma}c_{\alpha\gamma}
\label{11}~.
\end{equation}
By multiplying this expression on both sides by $\overline{|\alpha\rangle}$ and
summing over all these states, one gets
\begin{equation}
{\hat{\overline O}}_C|\Psi_{\gamma}\rangle
=\lambda_{\gamma}|\Psi_\gamma\rangle~,
\label{12}
\end{equation}
where $|\Psi_\gamma\rangle\equiv
\sum_\alpha c_{\alpha\gamma}{\overline{|\alpha\rangle}}$.
Therefore, the states $|\Psi_\gamma\rangle$ associated with the eigenvalues
$\lambda_\gamma$ are eigenstates of the operator ${\hat{\overline O}}_C$.

Let us now turn to the space $E$. Here, due to the orthonormality of the
states $|i)$, the identity operator is simply
\begin{equation}
{\hat I}_E=\sum_{i}|i)(i|
\label{15}~.
\end{equation}
Let ${\hat O}^{(nH)}_E$ be an operator acting within this space. Its action
on a state of $E$ is given by
\begin{equation}
{\hat O}^{(nH)}_E|l)=\sum_i |i)(i|{\hat O}^{(nH)}_E|l)
\label{104}~.
\end{equation}
By comparing Eqs.(\ref{101}) and (\ref{104}), one sees that, if
${\hat O}^{(nH)}_E$ is defined such that
\begin{equation}
(i|{\hat O}^{(nH)}_E |l)=\sum_{j}B(i,j)\langle j|{\hat O}_C|l\rangle
\label{17}~,
\end{equation}
its action on states of $E$ is formally identical to that of
${\hat{\overline O}}_C$ on the corresponding states of $C$.
As a result of that, if the state
\begin{equation}
|\Psi_\gamma\rangle =\sum_{\alpha}c_{\alpha\gamma}\overline{|\alpha\rangle}=
\sum_{i}(\sum_{\alpha}{\overline f}_{i\alpha}c_{\alpha\gamma})|i\rangle\equiv
\sum_{i}a_{i\gamma}|i\rangle
\label{13}
\end{equation}
is an eigenstate of ${\hat{\overline O}}_C$ corresponding to the eigenvalue
$\lambda_\gamma$, then, also the state
\begin{equation}
|\Psi_\gamma )=\sum_i a_{i\gamma}|i)
\label{105}
\end{equation}
is an eigenstate of ${\hat O}^{(nH)}_E$ with the same eigenvalue.
Therefore, ${\overline N}$ of the $N$
eigenvalues of ${\hat O}^{(nH)}_E$ in $E$ are the same as the eigenvalues
of ${\hat O}_C$ in $C$ and the associated eigenkets of ${\hat O}^{(nH)}_E$
are states ``simply'' corresponding to the
eigenkets of ${\hat{\overline O}}_C$ in $C$.
Eq.(\ref{17}) defines the image operator, non-Hermitian, of
${\hat O}_C$ in $E$.

Eq.(\ref{17}) recalls Eq.(23) of CS where, however, the
matrix $B(i,j)$  is replaced by the inverse of the overlap matrix. As we have
already seen, this can happen only in the case that ${\overline N}=N$.
In this case, characterized by the fact that
both ${\hat O}_C$ in $C$ and ${\hat O}^{(nH)}_E$ in $E$ would have the same
number of eigenvalues, each eigenstate $|\Psi_\gamma)$ of ${\hat O}^{(nH)}_E$
would be in a one-to-one correspondence with
an eigenstate $|\Psi_\gamma\rangle$ of ${\hat{\overline O}}_C$.
Therefore no ambiguities would exist about the ``physicality'' of all the
eigenstates of ${\hat O}^{(nH)}_E$.

The case ${\overline N}<N$, instead, namely the case under investigation,
appears more complicated. In this case, in fact, only ${\overline N}$
eigenstates of ${\hat O}^{(nH)}_E$ in $E$ can be in a one-to-one
correspondence with
eigenstates of ${\hat{\overline O}}_C$
in $C$ and so  have a physical meaning, while
the remaining $N-{\overline N}$ are unphysical and only a result of the
mapping procedure. In the following, we want to study
these eigenstates on the basis of the definition (\ref{17}) of the image
operator.

Let us first define, corresponding to each state (\ref{4}), the state
\begin{equation}
\overline{|\alpha)}=\sum_{i}{\overline f}_{i\alpha}|i)
\label{18}
\end{equation}
and let us call ${\overline E}$ the subspace of $E$ spanned by these states.
By noticing that
\begin{equation}
\overline{(\alpha}|\overline{\alpha')}=\slantfrac{1}{{\cal N}_\alpha}
\delta_{\alpha\alpha'}
\label{20}~,
\end{equation}
we conclude that this space is ${\overline N}$-dimensional. By multiplying
both sides of Eq.(\ref{17}) by the state $|i)$ and summing over all these
states, one obtains
\begin{equation}
\sum_{i}|i)(i|{\hat O}^{(nH)}_E|l)=\sum_{ij}|i)B(i,j)\langle j|{\hat
O}_C|l\rangle
\label{21}~.
\end{equation}
Moreover, by making use of Eqs.(\ref{4}), (\ref{10}) and (\ref{18}),
one obtains that
\begin{equation}
{\hat O}^{(nH)}_E|l)=\sum_{\alpha}\overline{|\alpha})
\langle\overline{\alpha|}{\hat O}_C|l\rangle
\label{22}
\end{equation}
and also
\begin{equation}
{\hat O}^{(nH)}_E\overline{|\beta)}=
\sum_{\alpha}\overline{|\alpha})
\langle\overline{\alpha|}{\hat O}_C\overline{|\beta\rangle}
\label{23}~.
\end{equation}
Let now $|\Psi_j)=\sum_{l}x_{lj}|l)$ be an eigenstate of ${\hat O}^{(nH)}_E$
with eigenvalue $\lambda_j$ and
$|\Psi_j\rangle =\sum_{l}x_{lj}|l\rangle$ its (non-zero) image state. Then
\begin{equation}
{\hat O}^{(nH)}_E|\Psi_j)
=\sum_{l}x_{lj}{\hat O}^{(nH)}_E|l)=\sum_{l}x_{lj}
\sum_{\alpha}\overline{|\alpha})
\langle\overline{\alpha|}{\hat O}_C|l\rangle =
\lambda_j\sum_{\alpha}\overline{|\alpha})\langle\overline{\alpha|}\Psi_j\rangle
\label{24}
\end{equation}
and since
\begin{equation}
{\hat O}^{(nH)}_E|\Psi_j)=\lambda_j|\Psi_j)
\label{25}~,
\end{equation}
one can conclude that, if $\lambda_j\neq 0$,
\begin{equation}
|\Psi_j)=\sum_{\alpha}\overline{|\alpha})\langle\overline{\alpha|}\Psi_j\rangle
\label{26}~.
\end{equation}
That is, in correspondence with an eigenvalue $\lambda_j\neq 0$, an eigenket
of the image operator ${\hat O}^{(nH)}_E$ must belong to ${\overline E}$.
But it has already been seen that ${\hat O}^{(nH)}_E$ has the
${\overline N}$ eigenkets (\ref{105}) and these can be rewritten as
\begin{equation}
|\Psi_\gamma)=\sum_{l}a_{l\gamma}|l)=\sum_{\alpha}c_{\alpha\gamma}
\overline{|\alpha)}
\label{27}~.
\end{equation}
These states belong to ${\overline E}$, are linearly independent (similarly
to the states (\ref{13})) and, this space being
${\overline N}$-dimensional, no extra eigenket linearly
independent from these can be accepted within this space.
As will be seen in subsect.3.3, even in the presence of degeneracies, all
the eigenstates of the non-Hermitian image ${\hat O}^{(nH)}_E$ in the space $E$
must be linearly independent.
Since, on the basis of Eq.(\ref{26}), a non-zero eigenvalue would force its
eigenstate to belong to $\overline E$ but this state could not be linearly
independent from the previous $\overline N$ eigenstates (\ref{27}), one can
only conclude that a $\lambda_j$ beyond those of these ${\overline N}$
eigenstates can not be different from zero. One finally notices that the case
of an image state $|\Psi_j\rangle$ with zero norm (we considered so far a
non-zero
state) corresponds to an eigenvalue $\lambda_j =0$, as it can be seen from
Eq.(42).

Eq.(\ref{17}) clearly shows that corresponding matrix elements of
${\hat O}_C$ and of its image operator ${\hat O}^{(nH)}_E$ are not equal.
In the following, however, we will show that a correspondence can be
established such that matrix elements in the quark and baryon spaces
can be preserved.
Due to the non-hermiticity of ${\hat O}^{(nH)}_E$, eigenbras and
eigenkets need not be dual vectors of one another. In order
to individuate the eigenbra corresponding to each eigenket (\ref{27}), let us
define the bras
\begin{equation}
\overline{\overline{(\alpha|}}=\sum_{i}\overline{\overline{f}}_{i\alpha}(i|
\label{28}~,
\end{equation}
where
\begin{equation}
\overline{\overline{f}}_{i\alpha}=\sqrt{{\cal N}_\alpha}f_{i\alpha}
\label{29}~.
\end{equation}
The space spanned by these states, $\overline{\overline{E}}$, is a
subspace of $E^{*}$, the dual space of $E$. Moreover, since
\begin{equation}
\overline{\overline{(\alpha}}|\overline{\overline{\alpha ')}}=
{\cal N}_{\alpha}\delta_{\alpha\alpha '}
\label{30}
\end{equation}
the dimension of $\overline{\overline{E}}$ is the same as that of
${\overline{E}}$, that is ${\overline{N}}$.

An important property of bras (\ref{28}) and kets (\ref{18}) is that
\begin{equation}
\overline{\overline{(\alpha}}|{\overline{\alpha ')}}=
\delta_{\alpha\alpha '}
\label{31}~.
\end{equation}
It follows from this and from Eq.(\ref{22}) that
\begin{equation}
\overline{\overline{(\alpha |}}{\hat O}^{(nH)}_E=\sum_{l}
{\overline{\langle\alpha |}}{\hat O}_C|l\rangle (l|
\label{32}~.
\end{equation}
By making use of this expression one can verify that the bra
\begin{equation}
(\overline{\overline{\Psi}}_\gamma|=\sum_{\alpha}c_{\alpha\gamma}
\overline{\overline{(\alpha |}}
\label{33}
\end{equation}
is an eigenbra of ${\hat O}^{(nH)}_E$ corresponding to the eigenvalue
$\lambda_\gamma$.

It also follows from (\ref{23}) and (\ref{31}) that
\begin{equation}
\overline{\overline{(\alpha |}}{\hat O}^{(nH)}_E{\overline{|\beta )}}=
{\overline{\langle\alpha |}}{\hat O}_C{\overline{|\beta\rangle}}
\label{34}~.
\end{equation}
Therefore, to any basis state ${\overline{|\alpha\rangle}}$
(\ref{4}) of $C$ one can associate a ket ${\overline{|\alpha )}}$ (\ref{18})
of ${\overline{E}}$ and a bra $\overline{\overline{(\alpha |}}$ (\ref{28})
of $\overline{\overline{E}}$ such that matrix elements of ${\hat O}_C$
between basis states of $C$ are equal to matrix elements of the image
${\hat O}^{(nH)}_E$ between the corresponding bra and ket of
$\overline{\overline{E}}$ and ${\overline{E}}$. Restricting the action of
${\hat O}^{(nH)}_E$ within ${\overline{E}}$ and
$\overline{\overline{E}}$, then, eliminates the unphysical zero
eigenvalues which emerge from the diagonalization of ${\hat O}^{(nH)}_E$
in the full $E$.

\subsection*{3.3 The image operator: its Hermitian form}

In order to derive the Hermitian form ${\hat O}^{(H)}_E$ of the image
operator defined in Eq.(\ref{17}), let us first introduce the operators
${\hat B}^{1/2}_E$ and ${\hat B}^{-1/2}_E$ such that
\begin{equation}
(i|{\hat B}^{1/2}_E|j)\equiv B^{1/2}(i,j)
\label{35}
\end{equation}
and
\begin{equation}
(i|{\hat B}^{-1/2}_E|j)\equiv B^{-1/2}(i,j)
\label{36}~,
\end{equation}
where $B^{1/2}(i,j)$ and $B^{-1/2}(i,j)$ are matrices real, symmetric
and such that
\begin{equation}
\sum_{i'}B^{1/2}(i,i')B^{1/2}(i',j)=B(i,j)
\label{37}
\end{equation}
and
\begin{equation}
\sum_{i'}B^{1/2}(i,i')B^{-1/2}(i',j)=\delta_{ij}
\label{38}~.
\end{equation}
${\hat B}^{1/2}_E$ and ${\hat B}^{-1/2}_E$ are, therefore, Hermitian
and such that
\begin{equation}
{\hat B}^{1/2}_E{\hat B}^{-1/2}_E={\hat B}^{-1/2}_E{\hat B}^{1/2}_E={\hat I}_E
\label{39}~.
\end{equation}
If $|\Psi_j)$ is an eigenket of ${\hat O}^{(nH)}_E$ associated with the
eigenvalue $\lambda_j$, one deduces that
\begin{equation}
{\hat B}^{-1/2}_E{\hat O}^{(nH)}_E{\hat B}^{1/2}_E{\hat B}^{-1/2}_E|\Psi_j)=
\lambda_j{\hat B}^{-1/2}_E|\Psi_j)
\label{40}~.
\end{equation}
By defining
\begin{equation}
{\hat O}^{(H)}_E\equiv
{\hat B}^{-1/2}_E{\hat O}^{(nH)}_E{\hat B}^{1/2}_E
\label{41}
\end{equation}
and
\begin{equation}
|\widetilde{\Psi}_j)\equiv {\hat B}^{-1/2}_E|\Psi_j)
\label{42}~,
\end{equation}
Eq.(\ref{40}) can be rewritten as
\begin{equation}
{\hat O}^{(H)}_E|\widetilde{\Psi}_j)=\lambda_j|\widetilde{\Psi}_j)
\label{43}~,
\end{equation}
namely $|\widetilde{\Psi}_j)$ is an eigenket of ${\hat O}^{(H)}_E$
associated with the same eigenvalue $\lambda_j$. It can be derived that
\begin{equation}
(i|{\hat O}^{(H)}_E|m)=\sum_{l,k}B^{1/2}(i,l)\langle l|{\hat O}_C|k\rangle
B^{1/2}(k,m)
\label{44}
\end{equation}
from which one deduces that ${\hat O}^{(H)}_E$ is indeed Hermitian. This
equation defines the image operator of ${\hat O}_C$ in $E$ in its Hermitian
form.

Eq. (\ref{42}) establishes a relation between eigenkets of ${\hat O}^{(nH)}_E$
and ${\hat O}^{(H)}_E$ corresponding to the same eigenvalue. The linear
independence of the first ones, stated in the previous subsection and not
guaranteed, in the presence of degeneracies,
for a non-Hermitian operator, is forced, as an effect
of this relation, by the linear independence of the eigenstates of
${\hat O}^{(H)}_E$ (obligatory for a Hermitian operator).

It has been seen in the previous subsection, Eq.(\ref{27}),
that eigenkets corresponding
to physical eigenvalues of ${\hat O}^{(nH)}_E$ are combinations of states
of ${\overline{E}}$. The state (\ref{42}) becomes in this case
\begin{equation}
|\widetilde{\Psi}_\gamma )=\sum_{\alpha}c_{\alpha\gamma}
{\hat B}^{-1/2}_E{\overline{|\alpha )}}=\sum_{\alpha}c_{\alpha\gamma}
\widetilde{|\alpha )}
\label{45}
\end{equation}
where
\begin{equation}
\widetilde{|\alpha )}\equiv {\hat B}^{-1/2}_E{\overline{|\alpha )}}=
\sum_j\{\sum_i\overline{f}_{i\alpha}B^{-1/2}(j,i)\}|j)
\label{46}~.
\end{equation}
One can verify that
\begin{equation}
\widetilde{(\alpha}|\widetilde{\alpha ')}=\delta_{\alpha\alpha'}
\label{47}
\end{equation}
and that
\begin{equation}
\widetilde{(\alpha |}{\hat O}^{(H)}_E\widetilde{|\alpha ')}=
\overline{\langle\alpha |}{\hat O}_C\overline{|\alpha '\rangle}
\label{48}.
\end{equation}
Therefore, if $\widetilde{E}$ is the $\overline{N}$-dimensional subspace
of $E$ spanned by the states $\{\widetilde{|1)}, \widetilde{|2)}, ...,
\widetilde{|\overline{N})}\}$ defined in Eq.(\ref{46}), one can say that to any
basis
state $\overline{|\alpha\rangle}$ (\ref{4}) of $C$ it is possible to associate
a state $\widetilde{|\alpha )}$ (\ref{46}) of $\widetilde{E}$ such that a
matrix element of ${\hat O}_C$ between basis states of $C$ is equal to the
matrix element of ${\hat O}^{(H)}_E$ between the corresponding states of
$\widetilde{E}$. The spectrum of ${\hat O}^{(H)}_E$ in $\widetilde{E}$
contains, then, only the physical part of the spectrum of ${\hat O}^{(H)}_E$
in $E$. The space $\widetilde{E}$ so defined individuates the physical
subspace of $E$.

It is interesting to discuss also the case of the eigenbra
$(\overline{\overline{\Psi}}_j|$ of ${\hat O}^{(nH)}_E$. Similarly to
Eq.(\ref{40}), one can write
\begin{equation}
(\overline{\overline{\Psi}}_j|{\hat B}^{1/2}_E{\hat B}^{-1/2}_E{\hat
O}^{(nH)}_E
{\hat B}^{1/2}_E=\lambda_j(\overline{\overline{\Psi}}_j|{\hat B}^{1/2}_E
\label{49}~,
\end{equation}
that is
\begin{equation}
(\widetilde{\widetilde{\Psi}}_j|{\hat O}^{(H)}_E=\lambda_j
(\widetilde{\widetilde{\Psi}}_j|
\label{50}~,
\end{equation}
where
\begin{equation}
(\widetilde{\widetilde{\Psi}}_j|\equiv (\overline{\overline{\Psi}}_j|
{\hat B}^{1/2}_E
\label{51}~.
\end{equation}
This state is the eigenbra of ${\hat O}^{(H)}_E$ corresponding to
the eigenket defined in Eq.(\ref{42}). More particularly, if
$\lambda_\gamma$ is a physical eigenvalue, the eigenbra corresponding
to the eigenket (\ref{45}) is
\begin{equation}
(\widetilde{\widetilde{\Psi}}_\gamma|=\sum_{\alpha}c_{\alpha\gamma}
\overline{\overline{(\alpha |}}{\hat B}^{1/2}_E
\label{52}~.
\end{equation}
It can be verified that
\begin{equation}
\overline{\overline{(\alpha |}}{\hat B}^{1/2}_E=\widetilde{(\alpha |}=
\overline{(\alpha |}{\hat B}^{-1/2}_E
\label{53}~,
\end{equation}
i.e. this bra is the dual vector of the ket $\widetilde{|\alpha )}$
defined in Eq.(\ref{46}). As expected, in this case, the eigenbra (\ref{52})
is simply the dual vector of the eigenket (\ref{45}). The two spaces
$\overline{E}$ and $\overline{\overline{E}}$ defined in the previous
subsection for kets and bras, respectively, are replaced here only
by the spaces $\widetilde{E}$ and its dual $\widetilde{E}^{*}$.

In conclusion, the operator ${\hat O}^{(H)}_E$ (\ref{41}) is a
Hermitian operator whose spectrum in $E$ is excatly the same as that of
${\hat O}^{(nH)}_E$ in $E$ and so contains (a) the $\overline{N}$
physical eigenvalues of ${\hat O}_C$ in $C$ and (b) the $N-\overline{N}$
unphysical zero eigenvalues. The physical subspace $\widetilde{E}$ of $E$
is spanned
by a set of $\overline{N}$ states in a one-to-one correspondence with the basis
states of $C$ and such that corresponding matrix elements of ${\hat O}^{(H)}_E$
and ${\hat O}_C$ are equal. The spectrum of ${\hat O}^{(H)}_E$ in
$\widetilde{E}$ coincides with that of ${\hat O}_C$ in $C$.

\section*{4. N-BODY STRUCTURE OF A BARYON OPERATOR: THE LIPKIN HAMILTONIAN}

In the previous section, we have derived the matrix elements defining the
baryon image of a quark operator. An important problem which we discuss
in this section is that related to the n-body structure of this baryon
operator. As an example, we will refer to the image of the Lipkin
Hamiltonian (\ref{2.3}).

The derivation of
the image operator treated in sect.3 has referred to the case
of the correspondence between the states (\ref{3.17}) and (\ref{3.22}),
namely, the case $\Omega =2$,
although, in principle, applicable also to larger values of $\Omega$.
The case $\Omega=1$, instead, the case of the correspondence between the states
(\ref{3.14}) and (\ref{3.21}), is particularly simple a case.
In this case, in fact, corresponding
states have the same overlaps so that the image
operator ${\hat H}_{E,1}$ of the Lipkin Hamiltonian is simply defined by the
equality
\begin{equation}
(\mu_1\mu_2\mu_3|{\hat H}_{E,1}|\mu_1'\mu_2'\mu_3')=
\langle\mu_1\mu_2\mu_3|{\hat H}_C|\mu_1'\mu_2'\mu_3'\rangle
\label{4.1}~.
\end{equation}
${\hat H}_{E,1}$ turns out to be the one-body Hermitian operator
\begin{eqnarray}
{\hat H}_{E,1}=
&&\slantfrac{3\Delta}{2}\sum_{p_1p_2p_3\sigma_2\sigma_3}
(f^{\dag}_{+p_1\sigma_2p_2\sigma_3p_3}f_{+p_1\sigma_2p_2\sigma_3p_3}-
f^{\dag}_{-p_1\sigma_2p_2\sigma_3p_3}f_{-p_1\sigma_2p_2\sigma_3p_3})\nonumber\\
&&-\slantfrac{12\chi_2}{\Omega}\sum_{p_1p_2p_3\sigma_3}
(f^{\dag}_{+p_1+p_2\sigma_3p_3}f_{-p_1-p_2\sigma_3p_3}+
f^{\dag}_{-p_1-p_2\sigma_3p_3}f_{+p_1+p_2\sigma_3p_3})\nonumber\\
&&-\slantfrac{36\chi_3}{\Omega^2}\sum_{p_1p_2p_3}
(f^{\dag}_{+p_1+p_2+p_3}f_{-p_1-p_2-p_3}+
f^{\dag}_{-p_1-p_2-p_3}f_{+p_1+p_2+p_3})
\label{4.2}~.
\end{eqnarray}
For $\Omega =2$, let us call
${\hat{\overline H}}_{E,2}$ the Hermitian image
of the Lipkin Hamiltonian which is defined by
Eq.(\ref{44}). One can find infinite combinations of
one-body plus two-body baryon operators satisfying this equation.
However, wishing the image Hamiltonian to be a good baryon image
for both $\Omega =1$ and $\Omega =2$ one is forced to take
\begin{equation}
{\hat{\overline H}}_{E,2}={\hat H}_{E,1}+{\hat H}_{E,2}~,
\label{4.3}
\end{equation}
where ${\hat H}_{E,1}$ has just been defined in Eq.(\ref{4.2}) and
${\hat H}_{E,2}$ is a two-body baryon operator defined by the matrix elements
\begin{equation}
(i|{\hat H}_{E,2}|j)=(i|{\hat{\overline H}}_{E,2}|j)-(i|{\hat H}_{E,1}|j)
\label{4.4}~.
\end{equation}
A similar procedure has to be extended to any $\Omega$ leading to the
general result that
\begin{equation}
{\hat{\overline H}}_{E,\Omega}={\hat H}_{E,1}+{\hat H}_{E,2}+...+
{\hat H}_{E,\Omega}
\label{4.5}~.
\end{equation}
As a general result, then, the image Hamiltonian is a baryon operator
containing up to $\Omega$-body terms
even if ${\hat H}_C$ is at most one-body. The presence of these
many-body terms
results from the need to simulate in the baryon space the complicated
underlying quark exchange dynamics.

\section*{5. Summary and conclusions}

We have discussed a mapping procedure
from a space of colorless three-quark clusters into a space of
elementary baryons and illustrated it within a three-color extension
of the Lipkin model. Special attention has been
addressed to the problem of the formation of unphysical states in the
mapped space.

The mechanism of the mapping
proposed has required, as first, to establish a correspondence between the
quark cluster space and the baryon space. Therefore, the baryon image
of a generic quark operator has been defined
in both its Hermitian and non-Hermitian forms and its spectrum analyzed.
As a general result, this spectrum
(equal in the two cases) has been found to consist
of a physical part containing the same eigenvalues of the quark
operator in the cluster space and an unphysical part consisting only of
zero eigenvalues. The last ones emerge as a product of the mapping
mechanism.

This derivation of the baryon image has not passed through
the preservation either of the commutation relations of the operators
or of the matrix elements in the quark and baryon  spaces. However, a
further correspondence has been established between quark and baryon
spaces such as to guarantee the equality of corresponding
matrix elements. A physical subspace of the baryon space
has been so defined.

We have examined the n-body structure of the image operator and considered,
as an example, the case of the Lipkin Hamiltonian.
In a correspondence involving $\Omega$ clusters, the need of up to
$\Omega$-body terms in the baryon operator, even in the case of
a one-body quark operator, has been discussed and the definition of these
terms provided.

With reference to the mapping procedure elaborated by PADF,
we notice that the baryon
Hamiltonian which has been constructed in that work, for $\Omega =2$,
has been found able to reproduce the spectrum of the Lipkin
Hamiltonian in the cluster space. In the cases examined, however,
unphysical eigenvalues have also appeared spread all over the spectrum,
in the Hermitian case, or pushed up in energy,
in the non-Hermitian case. This result clearly differs from that of the
present procedure, characterized by zero energy for all the unphysical
eigenstates and, therefore, by a better definite separation between
physical and unphysical eigenstates of the mapped Hamiltonian.

Although explicitly referring to the correspondence between systems of
3n fermions and n fermions, an important
aspect of the mapping procedure discussed in this paper consists in its
applicability to quite different scenarios like, for instance, the
correspondence between systems of 2n fermions and n bosons.
This can be clearly noticed in subsects. 3.2 and 3.3 where the formalism
of the procedure has been kept quite general just on purpose.
This ductility, together with its simplicity, makes this procedure available
for the most various applications.

\acknowledgments

The author wishes to thank S. Pittel, the discussions with whom have
stimulated this work. The author is also grateful to F. Catara for his
interest and G. Fonte and M.A. Nagarajan for useful discussions.


\begin{references}
\bibitem{isgur}
N. Isgur and G. Karl, Phys. Rev. {\bf D18}, 4187 (1978);
{Phys. Rev.} {\bf D19}, 2653 (1979);
{Phys. Rev.} {\bf D20}, 1191 (1979);
{Phys. Rev.} {\bf D21}, 3175 (1980);
M. Oka, K. Shimizu and Y. Yazaki, Nucl. Phys. {\bf A464}, 700 (1987);
K. Brauer, A. Faessler, F. Fernandez and K. Shimizu,
Nucl. Phys. {\bf A507}, 599 (1990).
\bibitem{shimi}
K. Shimizu, Rep. Prog. Phys. {\bf 52}, 1 (1989);
F. Myhrer and J. Wroldsen, Rev. Mod. Phys. {\bf 60}, 629 (1988).
\bibitem{yama}
Y. Yamauchi, R. Yamamoto and M. Wakamatsu, Nucl. Phys. {\bf A443}, 628 (1985);
P. Hoodbhoy, Nucl. Phys. {\bf A465}, 637 (1987);
S. Takeuchi, K. Shimizu and K. Yazaki, Nucl. Phys. {\bf A449}, 617 (1986).
\bibitem{klein}
A. Klein and E.R. Marshalek, Rev. Mod. Phys. {\bf 63}, 375 (1991).
\bibitem{nadja}
E.G. Nadjakov, J. Physics {\bf G16}, 1473 (1990).
\bibitem{padf}
S. Pittel, J.M. Arias, J. Dukelsky and A. Frank,
Phys. Rev. {\bf C50}, 423 (1994).
\bibitem{pedr}
S. Pittel, J. Engel, J. Dukelsky and P. Ring,
Phys. Lett. {\bf B247}, 185 (1990).
\bibitem{meyer}
J. Meyer, J. Math. Phys. {\bf 32}, 2142 (1991).
\bibitem{bely}
S.T. Belyaev and V.G. Zelevinsky, Nucl. Phys. {\bf 39}, 582 (1962).
\bibitem{maru}
T. Marumori, M. Yamamura and A. Tokunaga,
Prog. Theor. Phys. {\bf 31}, 1009 (1964);
T. Marumori, M. Yamamura, A. Tokunaga and A. Takada,
Prog. Theor. Phys. {\bf 32}, 726 (1964).
\bibitem{dyson}
F.J. Dyson, Phys. Rev. {\bf 102}, 1217 (1956).
\bibitem{petry}
H.R. Petry, H. Hofest\"adt, S. Merk, K. Bleuer, H. Bohr and K.S. Narain,
Phys. Lett. {\bf B159}, 363 (1985);
H. Hofest\"adt, S. Merk and H.R. Petry, Z. Phys. {\bf A326}, 391 (1987).
\bibitem{lipkin}
H.J. Lipkin, N. Meshkov and A. Glick, Nucl. Phys. {\bf 62}, 188 (1965).
\bibitem{cs}
F. Catara and M. Sambataro, Nucl Phys. {\bf A535}, 605 (1991).
\bibitem{oka}
M. Oka and K. Yazaki, Prog. Theor. Phys. {\bf 66}, 556 (1981).
\bibitem{cs2}
F. Catara and M. Sambataro, Phys. Rev. {\bf C46}, 754 (1992).
\end{references}
\end{document}